\documentclass [11pt]{article}
\usepackage{amsmath,amsthm,amsfonts,amscd,eucal,latexsym,amssymb}
\usepackage{epsfig}  
\def\sp{\hskip -5pt} 
\def\spa{\hskip -3pt}

\def\cH{{\ca H}}

\def\cO{{\ca O}}

\def\cA{{\ca A}}

\def\cK{{\ca K}}

\def\bC{{\mathbb C}}           

\def\bR{{\mathbb R}}

\def\bZ{{\mathbb Z}} 
 
\newsymbol\rest 1316         
 
\def\beq{\begin{eqnarray}}
\def\eeq{\end{eqnarray}}
\def\pa{\partial}
\def\at{\left(}               

\def\ct{\right)}              
\newcommand{\ca}[1]{{\cal #1}}         

\def\la{\lambda}

\def\si{\sigma}
\def\om{\omega}

\def\Ga{\Gamma}

\def\={\stackrel {\mbox{\scriptsize  def}} {=}} 
\newcounter{remark}[section]

\def\theremark{\thesection.\arabic{remark}}

\def\s #1 {\section{#1}}

\def\ssa #1 {\ifhmode{\par}\fi\refstepcounter{subsection}
  \noindent {\bf\thesubsection}. {\em #1}.\quad
  \addcontentsline{toc}{subsection}{\protect\numberline{\thesubsection} #1}%
  }

\def\ssb #1 {\ifhmode{\par}\fi\refstepcounter{subsection}
  \noindent {\bf\thesubsection.} {\bf #1.}\quad
  \addcontentsline{toc}{subsection}{\protect\numberline{\thesubsection} #1}%
  }

\def\remark {\ifhmode{\par}\fi\refstepcounter{remark}
  \noindent {\bf Remark \theremark}. \quad}

\newtheorem{teorema}{Theorem}[section]

\begin{document} 
%
%
 
\par 
\bigskip 
\LARGE 
\noindent 
{\bf ``Tunnelling''  black-hole radiation with  $\phi^3$ self-interaction: one-loop computation for Rindler Killing horizons } 
\bigskip 
\par 
\rm 
\normalsize 
 

\large
\noindent 
{\bf Giovanni Collini$^{1,a}$},  {\bf Valter Moretti$^{2,b}$},  {\bf Nicola Pinamonti$^{3,c}$} \\
\par
\small
\noindent$^1$Institut f\"ur Theoretische Physik, Br\"uderstr. 16 04103 Leipzig, Germany.
\smallskip

\noindent$^2$Dipartimento di Matematica, Universit\`a di Trento
 and  Istituto Nazionale di Fisica Nucleare -- Gruppo Collegato di Trento, via Sommarive 14  
I-38123 Povo (TN), Italy. \smallskip
\smallskip

\noindent $^3$Dipartimento di Matematica, Universit\`a di Genova,  
via Dodecaneso, 35
I-16146 Genova,  Italy. \smallskip
\smallskip

\noindent E-mail: 
$^a$giovanni.collini@itp.uni-leipzig.de,   $^b$moretti@science.unitn.it,  $^c$pinamont@dima.unige.it\\ 

 \normalsize

\par 
 
\rm\normalsize 

\rm\normalsize 

\rm\normalsize 
 
 
\par 
\bigskip 

\noindent 
\small 
{\bf Abstract}.  Tunnelling processes through black hole horizons have recently been investigated in the framework of WKB theory discovering interesting interplay  with the Hawking radiation.  A more precise  and general account of that phenomenon  has been subsequently given 
within the  framework of QFT in curved spacetime by two of the authors of the present paper.  
In particular,	it has been shown that,  in the limit of sharp localization on opposite sides of a Killing horizon, 
the quantum correlation functions of a scalar field appear to have thermal nature, and the tunnelling probability is proportional to $\exp\{-\beta_{Hawking} E\}$.
This local result is valid in every spacetime including a local Killing horizon, no field equation is necessary, while a suitable choice for the quantum state is relevant. Indeed,  the two-point function has to verify a short-distance condition weaker than the Hadamard one.
In this  paper we  consider a massive scalar quantum field with a $\phi^3$ self-interaction and  we investigate the issue whether or not 
the black hole radiation can be handled at perturbative  level,  including the renormalisation contributions.
We prove that, for the simplest model of  the Killing horizon generated by the boost in  Minkowski spacetime, and referring to Minkowski vacuum,   the tunnelling
 probability in the limit of sharp localization on opposite sides of the horizon preserves the thermal form proportional to $\exp\{-\beta_H E\}$ even taking the one-loop renormalisation  corrections into account.  A similar result is expected  to hold for the Unruh state in the Kruskal manifold, since that state is Hadamard and looks like Minkowski vacuum close to the  horizon.\\

\noindent {\bf Keywords}: Algebraic Quantum Field Theory,  Black hole radiation, Renormalisation, Rindler space.\\

\noindent{\bf MSC(2010) numbers}:  	81T15,  83C57, 81T20  
\normalsize
\bigskip

\s{Introduction }
\ssb{Hawking radiation as ``tunnelling process''}
In a couple of remarkable papers Parikh and Wilczek \cite{PW} and, independently, Volovik \cite{Vol} 
found that the {\em tunnelling probability}, $\Gamma_E$, of a particle with energy $E$ through a Schwarzschild Black Hole horizon is of thermal nature. 
Although their derivation is not theoretically clear, as we will discuss shortly,
 the characteristic Hawking temperature $T_H$ arises in their estimates.
%
%
%
This  result  would imply several physically notable consequences suggesting  a new viewpoint on the black hole radiation \cite{Hawking} issue. 
In particular, 
since no detection of radiation at future null infinity is exploited,
the mentioned result would suggest that the black hole radiation could be viewed as a completely {\em  local} phenomenon. 
More precisely,  only  the geometric structure in a 
neighborhood of a point on the horizon plays a role and thus no global black hole structure would  be necessary in this picture. 
Deliberately ignoring several conceptual issues (for the moment) and adopting  authors' point of view,  the mentioned  tunnelling  probability
is computed for {\em one particle} with energy $E$
moving between the events 
 $x=(t_1,r_1,\theta,\varphi)$ and $y=(t_2,r_2,\theta,\varphi)$, when these events are separated by the horizon and $x$ stays in the internal region.
The  understood quantization procedure is performed referring to the {\em Painlev\'e time} $t$ appearing in the corresponding 
explicit expression of the Schwarzschild metric.
The overall authors' idea is to  take advantage of  {\em WKB method} to approximate $\Ga_E$ and to study the leading order approximation 
for the  case of $x$ close to $y$, when the mass of the particle is negligible with respect to $E$:
$$
  \Ga_E  \sim  \lim_{y\to x}\left|e^{i\int_{r_1}^{r_2} p^{(E)}_r dr}\right|^2\;.
$$
However, a  difficulty appears: 
The integral in the exponent diverges.   In \cite{PW} this difficulty is turned out into an advantage by 
exploiting a suitable complex plane Feynman-like regularization. 
In this way an  {\em imaginary part} arises in the integral yielding:
\beq
\Ga_E \sim e^{-2 Im S_{reg}} \sim e^{-\beta_HE}\:, \quad \beta_H := 1/T_H\label{Ga}\:.
\eeq
The result has a natural interpretation in terms of a tunnelling process through the event horizon.
This, nowadays  very popular, result has been subsequently reproduced by various authors:
some unclear technical issues have been cleaned in \cite{APS,APGS};
the geometrical setting has been generalized even quite considerably, encompassing new
physically remarkable situations like {\em dynamical} black holes horizons;
other kind of particles have been considered and finally back reaction on the metric has been discussed
\cite{CaVol,fisici,vagenas,fisici2,mann,fisici3} (see \cite{fisici4} for a survey).

\noindent  However,  the presented machinery   remains  plagued by some unresolved problems analysed in \cite{MP}.
First of all,  the appearance of $T_H$ seems to be  suspiciously related with the choice of the adopted {\em complex-plane} regularization procedure. 
Furthermore  almost all  key tools, such as the  notions of {\em particle} (but also {\em time} and {\em energy}) are ambiguously defined in curved spacetime, due to the absence of  the Poincar\'e symmetry. However this is just one of the  problems.  Indeed, all mentioned papers  refer to ``a particle with energy $E$" and wavefunctions with definite energy which are localized etc. Instead,  particles are notoriously non-local concepts, and certainly an energy eigenstate can never be localized.  Energy itself is a non-local concept even in a flat spacetime.
Finally, despite  it is clearly suggested by the flavour of the  final result, it is by no means clear how the result is independent from the whole Black Hole geometry. This is because  
(\ref{Ga}) was obtained in \cite{PW} dealing with the Schwarzschild black hole manifold.\\

\ssb{The viewpoint of algebraic QFT in curved spacetime} 
 The rigorius framework of  {\em algebraic QFT in curved spacetime} was adopted in \cite{MP}
to clarify the physical meaning of Parikh-Wilczek's  result.  Let us review the outcome of that analysis referring  to \cite{Wald} for all geometric notions we employ. 
In a $4$-dimensional time-oriented smooth spacetime $M$ with Lorentzian metric $g$ having signature $-,+,+,+$, we consider an open set, ${\cal O} \subset M$, where
 a smooth vector field  $K$ exists satisfying the following requirements.
\begin{itemize}
  \setlength{\itemsep}{1pt}
  \setlength{\parskip}{0pt}
  \setlength{\parsep}{0pt}
\item[{\bf (a)}] $K$ is a Killing field for $g$ in $\cO$.
\item[{\bf (b)}] $\cO$ contains the {\bf local Killing horizon} ${\cal H}$ i.e. a $3$-submanifold invariant under the action of $K$ with $K^aK_a =0$ on ${\cal H}$.
\item[{\bf (c)}] The {\bf orbits} of $K$ in ${\cal O}$ are diffeomorphic to an {\bf open interval} $I$
and topologically ${\cal H}=I \times {\cal B}$ (${\cal B}$ being a $2$-dimensional cross section).
\item[{\bf (d)}]  The {\bf surface gravity} $\kappa \neq 0$ is  {\bf constant} on $\cH$.
{($\kappa$ is defined by $\nabla^a (K_bK^b)  = -2\kappa K^a$.)} 
\end{itemize}
\noindent We shall  make use of a standard {\em null coordinate system} $U,V,s$ adapted to $\cH$,  where $U \in I$ is the affine parameter of the  null geodesics forming $\cH$, 
$V$ is the affine parameter of the  null geodesics crossing $\cH$ once --  with the choice of the origin such that  $x\in \cH$ iff $V(x)=0$ --  and $s$ denotes a pair of coordinates over $\cal B$ where $U=0$. We refer to \cite{MP} for a precise definition.\\
As the computation will not depend on the geometry outside $\cO$,  the horizon may (smoothly) cease to exist outside $\cO$.
The requirement $\kappa=$ constant along $\cH$ means that the {\em  thermodynamic  equilibrium}
has been reached on $\cH$, since $\kappa = 2\pi T_H$. Notice that conditions (a)-(d) are quite general. For example they are satisfied around points of the {\em future horizon} of a {\em non-extremal}  black hole in the {\em Kerr-Newman family}, including {\em charged} and {\em rotating black holes}.
Furthermore, they are also valid both for ``realistic'' black holes produced by collapsed matter -- so that only the future horizon exists -- and even for {\em eternal  black holes} -- whose manifolds include {\em white hole} regions as in Kruskal spacetime.
Finally, our picture includes also situations where the collapse starts, reaches a sort of local equilibrium and it {\em stops} after a while, without giving rise to a complete BH structure.\\
Having discussed the geometric setup we pass now to present the quantum matter we consider. From now on $\cA$ is the unital  {\bf $*$-algebra} generated by  abstract  {\bf scalar  field operators}
 $\phi(f)$ with $f\in C^\infty_0(M)$ (the space of smooth complex and compactly supported functions on $M$) such that:
\begin{itemize}
  \setlength{\itemsep}{1pt}
  \setlength{\parskip}{0pt}
  \setlength{\parsep}{0pt}
\item[{\bf (R1)}]  $\phi(af +bf') = a\phi(f) + b\phi(f')$ if $a,b \in \bR$ and $f,f' \in C^\infty_0(M)$;
\item[{\bf (R2)}] $\phi(f)^*=\phi(\overline{f})$ for  $f\in C^\infty_0(M)$;
\item[{\bf (R3)}] $[\phi(f),\phi(f')]=0$ for causally disjoint $supp(f)$, $supp(f')$.
\end{itemize}	
Notice that, among these requirements {\em no field equation is assumed}. However, 
since we intend to compute the {\bf correlation function} $\om(\phi(f)\phi(f'))$ with respect to the state $\omega$ on $\cA$,
when $supp(f), supp(f') \subset \cO$ are ``very close'' to $\cH$, a condition on the state is necessary.  \\
 Supposing that $\cO$ is {\bf geodesically convex},  the signed squared geodesic distance of $x$ and $y$, $\sigma(x,y)$,  is thereon  well-defined, $t$ is any timelike coordinate  which increases towards the future and a standard $\epsilon \to 0^+$ prescription is assumed whenever indicated. With these notations, we assume that the integral kernel $\omega_2$ of $\om(\phi(f)\phi(f'))$ verifies the last hypothesis we need.
\begin{itemize}
 \setlength{\itemsep}{1pt}
  \setlength{\parskip}{0pt}
  \setlength{\parsep}{0pt}
\item[{\bf (R4)}] The short-distance behaviour holds:
 $$\om_2(p,p') :=    \frac{D(p,p')}{ \si_\epsilon(p,p')} + w_\epsilon(p,p')$$
where  $\si_\epsilon(p,p') := \si(p,p') +2i\epsilon (t_p-t_{p'}) + \epsilon^2$.\\
  $D$ is smooth and  a function $c : {\cal B} \to (0,+\infty)$ exists such that\footnote{In particular,  $c$ exists  if $D$ is both positive and invariant under the action of $K$ on $\cK$.
A stronger requirement on $D$ shows up in \cite{MP}, but actually  only the requirement above was exploited.} $D(p,p')= c(s_p)$ if $p,p'\in \cH$ and $s_p=s_{p'}$.\\ Finally $w_\epsilon$ is a  distribution {\em less singular} than $1/\sigma_\epsilon$.
\end{itemize}	
As in \cite{MP} we say that $w_\epsilon$ is {\em less singular} than $1/\sigma_\epsilon$ if
for every $\epsilon$, $w_\epsilon$ are measurable functions such that:\\
(i)  $w_\epsilon(p,p') \to w'(p,p')$, almost everywhere in $(p,p')$ as $\epsilon \to 0^+$
 for some function $w'$ and
$w_\epsilon$ is 
$\epsilon$-uniformly bounded by a locally $M^2$-integrable function;\\
(ii)  $w'(V,U, s, V ',U', s') \to  w''(U, s,U', s')$ almost everywhere in 
$(U, s,U', s')$
when $(V, V') \to (0, 0)$ for some function $w''$ on $\cH^2$ and $w'$  is $(V, V')$-uniformly
bounded by a locally $\cH^2$-integrable function.
\begin{remark}\label{remark} {\bf (1)}  
An example of $w_\epsilon$ less singular than $1/\sigma_\epsilon$  is, for every fixed $\delta>0$:
\beq 
w_\epsilon =   \frac{h_\epsilon(x,y)}{(\sigma_\epsilon(x,y))^{1-\delta}} +  f_\epsilon(x,y)\ln \si_\epsilon(x,y) + r_\epsilon(x,y)\quad \label{less}
\eeq  
where  for  fixed $\epsilon \in (0,\epsilon_0)$,    $f_\epsilon$, $r_\epsilon$, $h_\epsilon$ are measurable with $|f_\epsilon(x,y)|\leq K$ 
$|g_\epsilon(x,y)|\leq K'$ and $|h_\epsilon(x,y)|\leq K''$
 for  constants $K,K',K''$ and all $(\epsilon, x,y)\in (0,\epsilon_0)\times \cO\times \cO$ and, finally, 
 $f_\epsilon(x,y) \to f(x,y)$,  $r_\epsilon(x,y) \to r(x,y)$  and   $h_\epsilon(x,y) \to h(x,y)$ 
for $\epsilon \to 0^+$ if $\sigma(x,y) \neq 0$.
Above and throughout the cuts in the complex domain of $\ln z$ and $z^\alpha$ with $\alpha\not \in \bZ$ are supposed to 
stay along the negative real axis.	\\
We stress that  Gaussian {\em Hadamard states} for a linear Klein-Gordon field $\phi$  
satisfy the  requirement in (R4) with $w_\epsilon$ as in (\ref{less}) with  $h_\epsilon =0$.  \\
{\bf (2)} A requirement like (R4) was essentially assumed in \cite{FH}, dealing with  a linear scalar field, to prove that  this class of states produces  the black hole radiation at future null infinity for a  spacetime containing spherically symmetric collapsing matter giving rise to a black hole.\\
{\bf (3)} If we assume that  $\omega$ is Gaussian $f,f'$ are real,  and  $supp(f), supp(f')$ are separated by the horizon, then in the  Hilbert space of the 
GNS representation of  $\omega$, up to the normalization of the state,
$|\om(\phi(f)\phi(f'))|^2$ is nothing but  the transition probability
 of a pair of one-particle states $\phi(f) \Psi_\omega$ and $\phi(f')\Psi_\omega$. Here, $\Psi_\omega$ is the 
vacuum unit vector in the Fock-GNS representation of $\omega$,  localized on the opposite sides of the horizon.
\end{remark}\\

\noindent  We re-interpret the limit ``$x\to y$'' in Parikh-Wilczek picture as follows:
\beq 
\lim_{\la\to 0^+} \om(\phi(f_\la)\phi(f_\la')) = \lim_{\la\to 0^+} \om_2(f_\la,f_\la') \label{oo}
\eeq
where, as usual, on the right-hand side we denoted the distribution with the same symbol as its integral kernel. Furthermore, 
 the limit $\la \to 0^+$ {\em shrinks the supports of $f$ and $f'$ on $\cH$}.
Explicitly, making use of the previously mentioned null coordinate system $U,V,s$ adapted to $\cH$:
\beq
f_\la(V,U,x)=\frac{1}{\la} f\at \frac{V}{\la},U,s \ct \label{FL}\:.
\eeq
 To remove an infrared divergence arising in the computation of $\omega(\phi(f_\lambda)\phi(f'_\lambda))$
 as $\lambda \to 0^+$,
 we also assume that:
\beq \mbox{$f=\pa_V F$ and $f'=\pa_V F'$ for $F,F'\in C_0^\infty(\cO)$}\:. \label{fF}\eeq
Finally we need to specify the notions of {\em time} and {\em energy}, for the (locally) {\em external region} at least where $K$ is timelike. Therein $V \sim 
e^{-\kappa \tau}$.
The natural choice for time is the  {\em the parameter $\tau$ of the Killing field $K$}. The (locally) {\em internal region} is not stationary ($K$ is spacelike and
$V\sim -e^{\kappa \tau}$)
so no geometrically natural notion of time can exists there. Therefore we stick with $\tau$ in that region, too. The Fourier transform with respect to $\tau$ defines the {\em energy spectrum} with respect to 
the notion of energy $E$ associated with $\tau$.  We can state the main result of \cite{MP} (the  proof of the last statement is the same as that of (\ref{notunnell}) in \cite{MP}.)
\begin{teorema}\label{teo1}
Assuming that for the open set $\cO\subset M$ the hypotheses (a)-(d) hold and, referring to a state $\omega$ on $\cA$, that the requirements 
(R1)-(R4) hold true as well, for a certain positive mesure $\mu$ on ${\cal B}$, one has:
\beq
\lim_{\la\to 0^+}\omega(\phi(f_\la)\phi(f'_\la))=
\frac{\pi^2}{4}\int\limits_{\bR^2\times {\cal B}} {\int\limits_{\;-\infty}^\infty 
 \frac{\overline{\widehat{F}(E,U,x)}  \widehat{F'}(E,U',x)}{\sinh(\beta_H E/2)}  EdE}\: dU dU' 
c(x)d\mu(x) \:,\label{tunnel} 
\eeq
where $f$ and $f'$ are real, have  supports separated by the horizon, verify (\ref{FL}),(\ref{fF}) and $\widehat{F}(E,U,x)$, $\widehat{F'}(E,U,x)$ denotes
the $\tau$-Fourier transform  of ${F}(e^{-\kappa \tau},U,x)$ and ${F'}(-e^{-\kappa \tau},U,x)$ respectively.
 For  wave packets sharply concentrated around a large value $E_0>0$ of the energy,  (\ref{tunnel}) yields:
\beq 
\lim_{\la\to 0}|\omega(\phi(f_\la)\phi(f'_\la))|^2 \sim C  E_0^2 \: e^{-\beta_H E_0}\:,\label{estim}
\eeq
where $C$ does not depend on $\beta_H$.\\
If both the supports of the real functions $f$ and $f'$ stay in the external region a proper Bose spectrum arises:
\beq
\lim_{\la\to 0^+}\omega(\phi(f_\la)\phi(f'_\la))=\frac{\pi^2}{8}
\int\limits_{\bR^2\times {\cal B}} {\int\limits_{\;-\infty}^\infty 
 \frac{\overline{\widehat{F}(E,U,x)}  \widehat{F'}(E,U',x)}{1-e^{-\beta_H E}}  EdE}\: dU dU' 
c(x)d\mu(x) \:.\label{notunnell} 
\eeq
An identity like (\ref{notunnell}), with
$\overline{\widehat{F}}\widehat{F'}$
replaced by  $\widehat{F}\overline{\widehat{F'}}$ in the integrand,
 holds 
for real $f,f'$ both supported in the internal region.
\end{teorema}\label{teo}

\noindent 
To estimate the leading order for the {\em transition probability} from one side of the horizon to the other one, we normalize
dividing both sides of (\ref{estim}) by the product of  squared norms 
$\omega(\phi(f_\lambda)\phi(f_\lambda)) = ||\phi(f_\lambda)\Psi_\omega||^2$,
$\omega(\phi(f'_\lambda)\phi(f'_\lambda)) = ||\phi(f'_\lambda)\Psi_\omega||^2$ and then we take the limit.
Proceeding in this way we obtain a result similar to the right hand side of  (\ref{estim})  but with a different constant $C'$ which takes the normalization of the vectors into account. 
Nevertheless, it follows from the estimate of $\omega(\phi(f_\lambda)\phi(f_\lambda))$ and $\omega(\phi(f'_\lambda)\phi(f'_\lambda))$ given by 
 (\ref{notunnell}) with  $f=f'$ and for $\beta_H E_0 >\spa> 1$  and form the last statement of Theorem \ref{teo1} that
 again, $C'$ does not depend on $\beta_H$ for packets sharply concentrated around a large value $E_0>0$.
In this way, adopting the viewpoint of algebraic QFT in curved spacetime,
Parikh, Wilczek and Volovik's result acquires a  precise and  rigorous 
meaning,  though the tunnelling interpretation does not take place strictly  speaking.
 As our computation  is completely {\em local in space and time}, it strongly supports the idea that the {\em Hawking radiation is (also) a local phenomenon}, independent from the existence of a whole black hole. 
The result is  {\em  independent form the state} of the quantum field, provided it belongs to a large class including the physically significant {\em Hadamard states}. 
That class of states  enjoys physically fundamental properties
in developing linear QFT in curved spacetime and  in the  semiclassical treatment of quantum gravity \cite{Wald2}. Referring to those states,
the back-reaction on the metric can be computed because they admit a well-defined  stress energy tensor \cite{Mo}. 
Moreover,  considering interacting quantum fields  adopting a perturbative approach, a generally locally covariant  renormalisation procedure can be successfully implemented referring to Hadamard  states \cite{BF, Howa01, Howa02, BDF}. That procedure is similar and generalizes  the standard renormalisation machinery   in flat spacetime  developed with respect to  the standard free Poincar\'e invariant  vacuum of the free theory.\\
A final remark concerning the value of $T_H$ in our local picture is necessary.  Without fixing the value of $K$ at some point, a constant rescaling ambiguity remains in the definition of $K$, affecting the value of $T_H= \kappa/(2\pi)$. In a black-hole manifold which is  asymptotically flat  this ambiguity is removed assuming that $K$ defines  the Minkowski standard time far away from the horizon.  In the general case, even if the mentioned ambiguity exists,  the {\em local temperature} $T_H(x)$  measuraed  by a thermometer  at rest with $K$ is however well defined.  Indeed, by definition  $T_H(x) := T_H/ \sqrt{-K_a(x)K^a(x)})$ where the red-shift Tolman factor  \cite{Wald2} $(-K_a(x)K^a(x))^{-1/2}$ appears.
It is obvious from the definition of $\kappa$ that $T_H(x)$ is fixed if constantly rescaling $K$ by a factor $c>0$. Indeed, the said rescaling produces $\kappa \to \kappa' = c\kappa$
and thus
 $T_H(x) \to T'_H(x) = cT_H/ \sqrt{-cK_a(x)cK^a(x)}) = T_H/ \sqrt{-K_a(x)K^a(x)}) = T_H(x)$.\\

 \ssb{Motivation and main result of this work} 
The result of \cite{MP} does not  depend on any field equation, but only on the short distance behavior (R4) of the two-point function of the considered state.
Thus, it must be true even considering {\em interacting quantum fields} provided a suitable scaling limit of the two-point function holds \cite{FH-scaling,Buchholz}. 
 It is however far from obvious that (R4)  also  holds when treating QFT perturbatively,  taking the {\em renormalisation} corrections into account and starting for a state of the free theory verifying (R4).\\
In this work,  we will focus on the simplest model given by the ${\cal L}_I = \frac{g}{3!}\phi^3$ self-interaction in Minkowski spacetime, referring to the Killing horizon generated by a boost vector field. The main idea developed within this paper  is, in fact, to compute the renormalized two-point function at one loop approximation for the Poincar\'e invariant vacuum state
(that obviously verifies (R4) when radiative corrections are disregarded) and to check whether it satisfies the requirement (R4);  (R1)-(R3) being automatically true for a real quantum scalar field. If it is the case, taking (\ref{oo}) into account with 
$\omega_2$ given by the one-loop renormalized two-point function,  Theorem \ref{teo} authorizes  one  to  conclude that
the Hawking radiation viewed as a local (``tunnelling'') phenomenon survives the introduction of a $\frac{g}{3!}\phi^3$ self-interaction, at one loop at least. \\
The main result of this work, explicitly stated  at the end of section \ref{lastsect},  is that the requirement (R4) is actually fulfilled by the (one-loop) renormalized two-point function, so that the local Hawking radiation appears even taking the self-interaction into account at one-loop.

\s{The simplest interacting case: ${\cal L}_I = \frac{g}{3!}\phi^3$ in Rindler spacetime}
\ssb{Comparison with more physical cases}
Before going on with computations, let us  briefly discuss why we expect that this simple Minkowskian model makes sense from a physical viewpoint in comparison with the analogous situation for a Schwarzschild black hole.  In the latter case, dealing with the Kruskal manifold, the physically interesting state is the celebrated  {\em Unruh state}, since it is the natural state where Hawking radiation is detected at future null infinity. An explicit rigorous construction of that state has recently   been established in \cite{DMP} where, in particular,  the state has been shown to be of Hadamard type in the model of a real black hole 
spacetime  made of the union of the black hole  region and the right Schwarzschild wedge (regions  I and III in  Fig.5.1 of \cite{Wald2}) of the Kruskal manifold.  So, that state  verifies the requirement (R4) in a neighborhood of the future (right) Killing event horizon. Moreover, as it was already known from heuristic constructions, that state looks like Minkowski vacuum as soon as one approaches the Killing horizon. On the other hand the very geometry of Kruskal manifold  locally approximates Minkowski one as soon as one approaches the Killing horizon.   The differences appear far from the Killing horizon. In the Schwarzshild manifold
  the Killing field defining the Killing horizon becomes the Minkowski time vector giving rise to the natural notion of the energy far away from the black hole. 
Instead, in Minkowski space, the Killing field defining the Rindler horizon does not approach the Minkowski time vector far from the horizon. However, it might  not matter since we are interested in what happens close to the horizon. Our approximation allows us to exploit the relatively simpler version of renormalisation procedure in flat spacetime than  the generally locally covariant version
 in curved spacetime \cite{Howa01,Howa02,BDF}.  A  large number of counter terms arising from  the curvatures will be completely neglected in our elementary model.
Certainly, a quantum state and the renormalisation procedure are non-local concepts,  so there is no guarantee to automatically extend a positive result found in Minkowski space for the Poincar\'e invariant state  to the Kruskal manifold and referring to the Unruh state, although the only local structure of the two-point function seems to be relevant.  However, if the black hole radiation (viewed as a ``tunnelling probability'') did not survive the introduction of a self-interaction in Minkowski spacetime, it very unlikely  would do in curved spacetime.\\

\ssb{Computation} In the following  $\phi$ and $\phi_0$ respectively denote the renormalized and free (massive  Klein-Gordon) quantum field. The same convention is exploited for Minkowski vacua, $\Psi$  and  $\Psi_0$ respectively.
The Gell-Mann Low formula for time-ordered two-point functions holds:
\beq
\langle \Psi , T[\phi(x)\phi(y)]\Psi\rangle  = \frac{\langle \Psi_0 , T[\phi_0(x)\phi_0(y) S(g)] \Psi_0\rangle }{\langle \Psi_0, TS(g) \Psi_0 \rangle} \label{GL}
\eeq
where formally:
\beq
S(g) = I +i  \int_M \spa \frac{g(u)}{3!} :\spa\phi_0^3\spa:\spa(u) d^4u - \frac{1}{2!} \int_M\spa \int_M \spa \frac{g(u)}{3!}  \frac{g(u')}{3!} :\spa\phi_0^3\spa:\spa(u)\spa:\spa\phi_0^3\spa:\spa(u') d^4u d^4u'+ \cdots \label{Sg}\:.
\eeq
Above we assume that the  Wick monomials, denoted by $:\cdot :$ are those defined with respect to $\Psi_0$ (their expectation values vanish on $\Psi_0$) and
the function $g \in C_0^\infty(M)$, attaining constantly the value $g_0$ in a bounded region, has to be switched  to an everywhere  constant function 
at the end of computations. This is done to remove infrared divergences. 
Later, we will extract the two-point function from the {\em time-ordered} two-point function.
 However this would not be truly necessary for,  if $x$ does not belong to the causal past of  $y$:
\beq 
\omega_2(x,y) := \langle \Psi ,\phi(x)\phi(y) \Psi\rangle  = \langle \Psi , T[\phi(x)\phi(y)] \Psi\rangle\label{fundamental}\:.
\eeq
This is the very situation  when $x$ and $y$ are separated by a Killing horizon and $x$ stays in the non-static region while $y$ stays in the static one (the right Rindler wedge in our case).\\
In the following, we will compute  the explicit expression of the one-loop renormalized  two-point function in the position domain, rather than 
in the momentum space, because we intend to check whether or not the requirement (R4) still holds taking the radiative corrections into account. 
Here, we are looking for a very precise expression avoiding formal computations  based, for instance, on divergent series or large momentum approximations.   
It is of course already known that the singularities of two-point function for the theory with radiative corrections 
are for points which are light-like related. 
However, since such a distribution does not solve the Klein Gordon equation, the knowledge of the form of its wave front set is not sufficient to apply the result of Radzikowski \cite{Radzikowski} to conclude that its small distance behavior is of the form (R4).

\begin{figure}
\centering
   \scalebox{0.4}{\begin{picture}(500,180)(0,0) 
       \put(-30,0){\includegraphics[height=10cm]{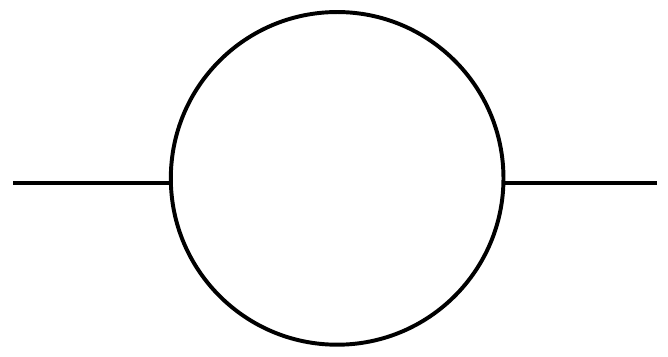}}
   \scalebox{2.5}{\put(-20,50){$x$}}
   \scalebox{2.5}{\put(200,50){$y$}}
   \scalebox{2.5}{\put(50,50){$u$}}
 \scalebox{2.5}{\put(130,50){$u'$}}
       \end{picture}}
     \caption{The diagram corresponding to $\tau(x,y,u,u')$ in Eq.(\ref{tau4}) }
\end{figure}
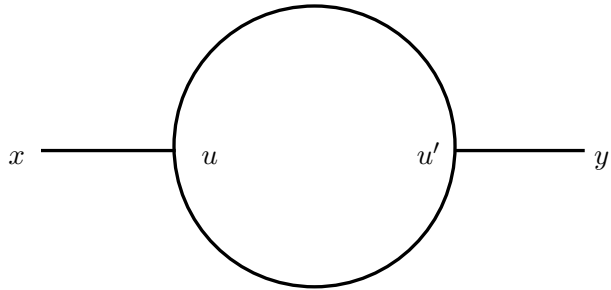

Looking at the right hand side of (\ref{Sg}) one realizes that the first non-trivial contribution to the right-hand side of  (\ref{GL}) comes from the double integral, that is the diagram in Figure 1, because the previous term yields a vanishing contribution.  (In curved spacetime  adopting the general locally covariant notion of Wick polynomial of \cite{Howa01} or even in Minkowski spacetime  referring the normal ordering to a state different form Minkowski one, also the second  term in the right-hand side of (\ref{Sg}) would give a non-trivial  contribution.)
 The relevant  $4$-point function is therefore  the distribution on $C_0^\infty(M^4)$ corresponding to the diagram in figure 1:
\beq \tau(x,y,u,u') := 
\langle \Psi_0,T[\phi_0(x)\phi_0(y) \spa:\spa\phi_0^3\spa:\spa(u)\spa:\spa\phi_0^3\spa:\spa(u') ] \Psi_0 \rangle\:.\label{tau4}\eeq
As is well-known this is not a well-behaved distribution since  it is well-defined only for test functions whose supports do not intersect 
the diagonals of the product $M\times M \times M \times M$. The extension procedure and the classification of the arising ambiguities is nothing but 
the Epstein-Glaser version of the renormalisation procedure \cite{EG}, that has been generalized in curved spacetime to a generally locally covariant procedure \cite{BF,Howa01,Howa02,BDF}. The ambiguities, i.e.,  the finite-renormalisation counter terms,  are classified   imposing constraints concerning, covariance, causal factorisation,  scaling behaviour  and polynomial dependence on the mass and the inverse metric. Dropping terms vanishing in the adiabatic limit,
the only counter term for the above $\tau(x,y,u,u')$ is proportional to
$$\delta \tau(x,y,u,u')  := \delta(u,u')\langle \Psi_0,T[:\spa\phi^2_0\spa:\spa(u')\phi_0(x) \phi_0(y) ] \Psi_0 \rangle \:.$$
If $G_F= i \langle \Psi_0 , T[\phi_0(x)\phi_0(y)]\Psi_0\rangle $ is the free  Feynman propagator  we therefore have:
$$\langle \Psi , T[\phi(x)\phi(y)]\Psi\rangle = -iG_F(x,y)$$
\beq
+ A \spa  \int_{M}\sp\sp g(u)^2 \: G_F(x,u)  G_F(u,y) d^4u
- \frac{1}{2!}\spa\int_{M^2}\sp\sp \spa g(u)g(u') \: G_F(x,u) G^{2(ext)}_F(u,u') G_F(z',y) d^4ud^4u'+\cdots \label{sum}
\eeq
where  $A$ is a finite renormalisation constant.
The first convolution in right hand side is well defined, as it follows by direct inspection in flat spacetime or  in view of general theorems on microlocal analysis in curved spacetime  \cite{Howa01,Howa02}).  $G_F^2$ is 
well defined as a distribution only on $C^\infty_0(M\times M \setminus \Delta_2)$, where $\Delta_2:= \{(x,x)\:|\: x\in M\}$,
again it follows either  by direct inspection in flat spacetime or  in view of general theorems on microlocal analysis in  curved spacetime. Thus, in the second convolution in (\ref{sum}), it has been replaced 
for  an extension $G^{2(ext)}_F$ acting on the whole $C^\infty_0(M\times M)$.
Throughout we use the 
conventions of \cite{Stro} about Fourier transform ($f(x) = (2\pi)^{-2}\int e^{ik_\mu x^\mu}\hat{f}(k) d^4k$)  and propagators
and decompose four vectors as  
 $s=(s^0, {\bf s})$ with $s^2:= -(s^0)^2 + {\bf s}\cdot {\bf s}$.
Passing to the Fourier transform of distributions, a well-known extension of $G_F^2$ is given by:
		\begin{eqnarray}
		\widehat{G^{2(ext)}_F} (k) &=&\frac{1}{(2\pi)^6} \int_{\bR^4} \left[\frac{1}{p^2 + m^2 -i \epsilon} \frac{1}{(p+k)^2 +m^2 -i \epsilon}  -  \frac{1}{(p^2 + m^2 - i \epsilon)^2}\right] d^4p \nonumber 
		\end{eqnarray}
Above and henceforth,  a distributional $\epsilon\to 0$ limit is {\em implicit}.
	Making  use of the standard  Feynman parameters procedure,
	after a lengthy integral computation we obtain:
		\begin{eqnarray}
			\widehat{G^{2(ext)}_F} (k)  
			&=&\frac{i}{2(2\pi)^4}\left[ -1 + \sqrt{1 + \frac{4m^2 - i \epsilon}{k^2}} \coth^{-1} \left( \sqrt{1 + \frac{4m^2 - i \epsilon}{k^2}} \right) \right]\:. \nonumber
		\end{eqnarray}
Inserting the result  in (\ref{sum}), exploiting the convolution theorem and performing the adiabatic limit, since everything is well defined, so that $g$ is constant, we find:
\begin{gather}
\frac{-1}{2!}\int_{M^2}\sp\sp \spa g^2 \: G_F(x,u) G^{2(ext)}_F(u,u') G_F(u',y) d^4ud^4u' = \nonumber \\
  \frac{-ig^2 }{4(2\pi)^6}\int_{\bR^4}  \frac{e^{i k_\mu (x-y)^\mu}}{(k^2 + m^2 - i\epsilon)^2}  \left[ -1 + \sqrt{1 + \frac{4m^2 - i \epsilon}{k^2}} \right.
			 \left. \coth^{-1} \left( \sqrt{1 + \frac{4m^2 - i \epsilon}{k^2}} \right) \right] d^4k\:. \label{inter}
\end{gather}
The integration in $k^0$ can be computed before that in ${\bf k}$ extending $k^0$ to a complex variable $z$. It is done by completing the integration along the real line into a closed contour with an arch at infinity in either the lower or the upper half-plane, depending on the sign of $(t_x-t_y)$,  taking advantage of the  residue technology as is well known. 
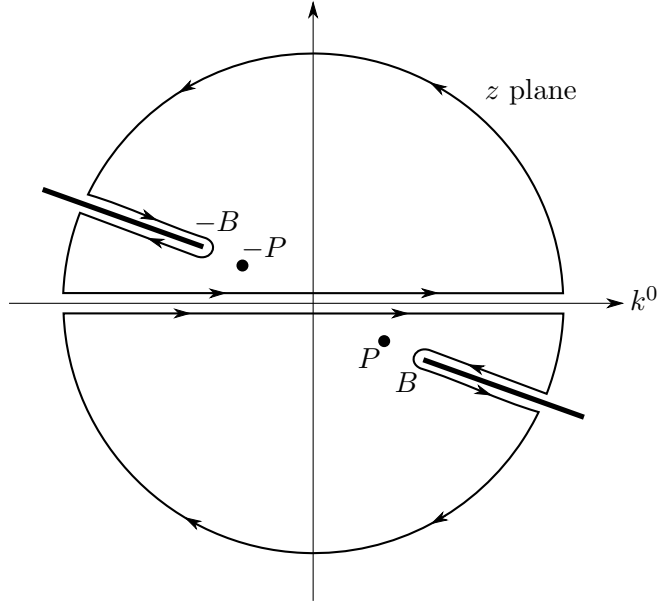
\begin{figure}
\centering
     \begin{picture}(160,180)(0,0) 
       \put(-30,0){\includegraphics[height=8cm]{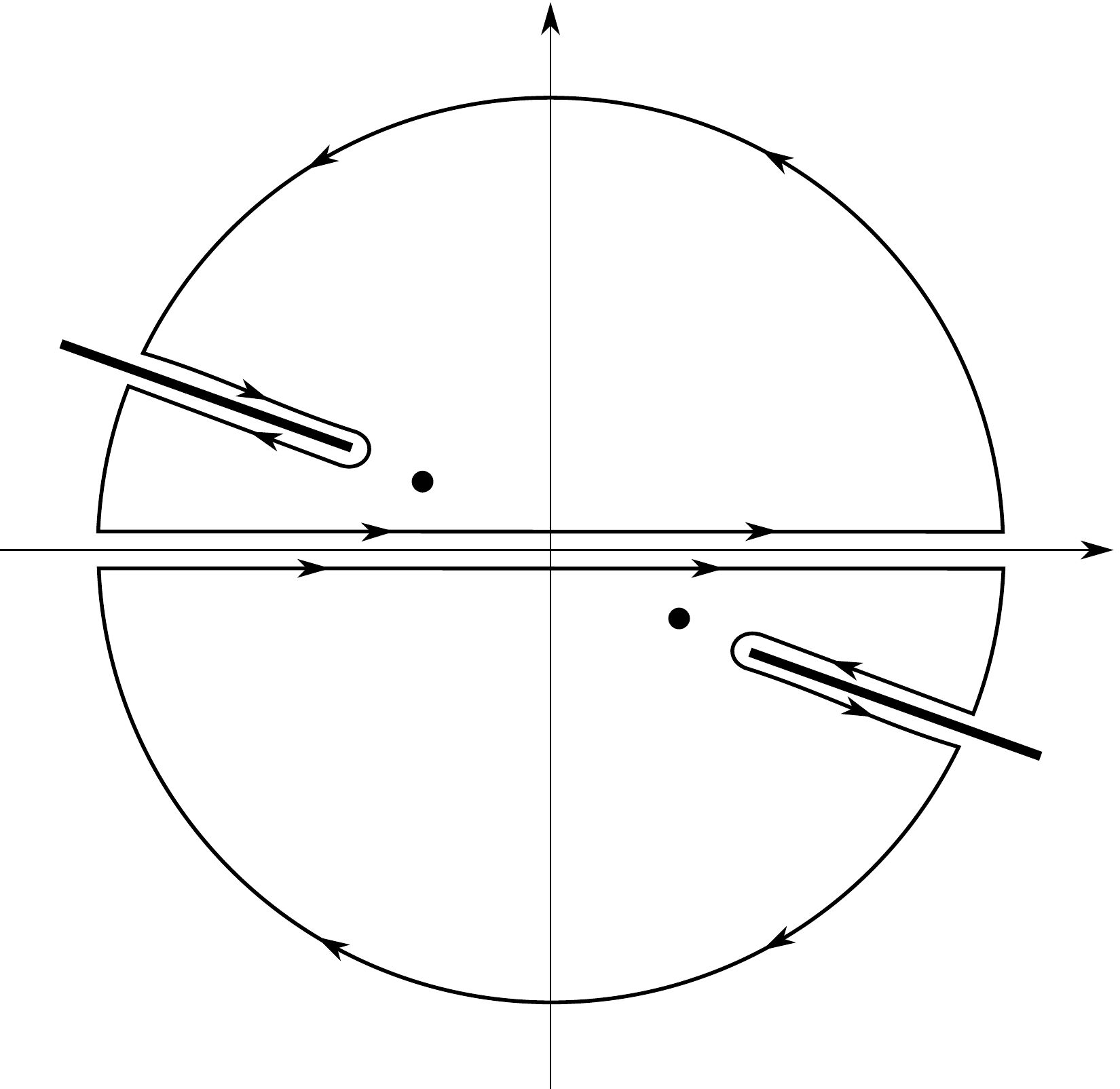}}
  \put(102,88){$P$}
  \put(150,190){$z$ plane}
       \put(205,110){$k^0$}
       \put(116,79){$B$}
       \put(58,130){$-P$}
       \put(40,140){$-B$}
       \end{picture}
     \caption{The picture emphasizes the two poles at $z=\pm P =\pm\sqrt{{\bf k}^2 + m^2 -i\epsilon}$ and the two branch cuts (starting at $z=\pm B=\pm\sqrt{{\bf k}^2 + 4m^2 -i\epsilon}$) relevant for the 
     complex extension $z$ of $k^0\in \bR$.
     It represents also the two contours on which the $z$ integration is taken 
     for positive (lower contour) and negative (upper contour) $t_x-t_y$. 
     }
\end{figure}
The analytic continuation (in the variable $k^0$) of the integrand of (\ref{inter}) gives rise to a couple of poles at $\pm\sqrt{{\bf k}^2 + m^2 -i\epsilon}$ respectively.
However, although no problems arise with the decay rapidity of the considered functions on the portion of the contour at  infinity,   
 a closer scrutiny of the function in square brackets in (\ref{inter}) added to $-1$ reveals  the appearance of a pair of  logarithmic branch cuts. One, 
relevant for $t_x-t_y >0$, completely stays
 in the lower half-plane  starting from $\sqrt{{\bf k}^2 + 4m^2 -i\epsilon}$ and joining $\infty$;  the other, relevant for $t_x-t_y<0$, completely stays  in the upper half plane starting from 
 $-\sqrt{{\bf k}^2 + 4m^2 -i\epsilon}$ and joining $\infty$. So the contributions of these branch cuts have to be taken into account.
Making explicit the contribution of the poles, introducing an $\epsilon$-prescription in the spacetime representation necessary to interchange some integrations, and indicating by $C(x,y)$ the contribution due of the branch cuts, the result is: 
		\begin{eqnarray}
&\;&\frac{-g^2}{2!}\int_{M^2}\sp\sp  G_F(x,u) G^{2(ext)}_F(u,u') G_F(u',y) d^4ud^4u' = 
\nonumber\\
			&=& \frac{g^2}{4(2\pi)^2} \left( \frac{\pi \sqrt{3}}{3} -\frac{1}{2}\right) K_0\left( m \sqrt{\sigma(x,y) + 2i |t_x-t_y|\epsilon +\epsilon^2}\right)  \nonumber \\
			&\;& 
			-\frac{g^2}{4(2\pi)^2} \left( \frac{1}{2}-\frac{\pi}{3\sqrt{3}} \right) \frac{K_1\left( m \sqrt{\sigma(x,y) + 2i |t_x-t_y|\epsilon +\epsilon^2}\right) }{ m \sqrt{\sigma(x,y) + 2i |t_x-t_y|\epsilon +\epsilon^2}} + C(x,y)  \label{sum2}
		\end{eqnarray}
		where $K_\nu$ are the well-known modified Bessel functions of the second kind. We have used formulas 3.914 (10) and (9) of \cite{Grad}.
The term proportional to the undetermined constant $A$ in (\ref{sum}) can be analogously computed and the only result is to change the factor 
in front of $K_0$ by an unknown constant. The  term  in (\ref{sum2}) immediately before $C(x,y)$,  if replacing  the coefficient in front of $K_1$ by $i \frac{m^2}{(2\pi)^2}$
is nothing but $G_F(x,y)$.  Putting all together,
 (\ref{sum}) yields the result:
$$\langle \Psi, T[\phi(x)\phi(y)] \Psi\rangle =A K_0\left( m \sqrt{\sigma(x,y) + 2i |t_x-t_y|\epsilon +\epsilon^2}\right) 
$$ \beq  +\left[  \frac{m^2}{(2\pi)^2}
 -  \frac{g^2}{4(2\pi)^2} \left( \frac{1}{2}-\frac{\pi}{3\sqrt{3}} \right) \right]
\frac{K_1\left( m \sqrt{\sigma(x,y) + 2i |t_x-t_y|\epsilon +\epsilon^2}\right) }{ m \sqrt{\sigma(x,y) + 2i |t_x-t_y|\epsilon +\epsilon^2}}
+ C(x,y) + \cdots \label{almostf}\eeq
It remains to evaluate $C(x,y)$. Using the very definition of $\coth^{-1}$ and the well-known fact that $\ln(-|x| + i\delta) - \ln(-|x| - i\delta) \to 2i\pi $ for $\delta \to 0^+$
in evaluating the integral along curve surrounding a branch cut  of the function added to $-1$  in square brackets in (\ref{inter}),
one finds:
$$C(x,y) = \frac{g^2}{4(2\pi)^4|{\bf x}-{\bf y}|} \int_0^{+\infty} \spa\spa \sp \sp d|{\bf k}| |{\bf k}|  \int_\gamma dz \frac{e^{-i z |t_x-t_y|} \sin( |{\bf x}-{\bf y}| |{\bf k}|) }{(z^2 - {\bf k}^2 -m^2 +i\epsilon)^2} \sqrt{1+ \frac{4m^2-i\epsilon}{{\bf k}^2 -z^2}}
$$
where $\gamma$  is the lower branch cut. However, since the integrand of the $z$ integration  is holomorphic in the lower half plane (barring a branch cut from $z=|{\bf k}|$ to $z=\sqrt{{\bf k}^2 + 4m^2 -i\epsilon}$) and it decreases rapidly, $\gamma$ can be deformed without affecting the value of the integral, provided  the path keeps joining $\sqrt{{\bf k}^2 + 4m^2 -i\epsilon}$ and $\infty$ (and avoids  the cut). For convenience we therefore  assume $\gamma$ to be of the form
  $z(s) = \sqrt{{\bf k}^2 + 4m^2 -i\epsilon +s^2}$ with $s \in [0, +\infty)$. The integration in $ds$ can be evaluated after computing that in $d|{\bf k}|$ obtaining:
\beq C(x,y)=  {\cal K} \left(\sigma(x,y)+2i|t_x-t_y|\epsilon +\epsilon^2\right)\label{cK}\:,\eeq
where 
$M^2 = s^2 + 4m^2$ and:
\beq
{\cal K}(u) := \frac{g^2}{4(2\pi)^4} \frac{1}{\sqrt{u}}\int_{2m}^{+\infty}\sp\sp\spa dM \frac{\sqrt{M^2 -4m^2}}{(M^2-m^2)^2}M
 K_1\left(M\sqrt{u}\right) \;.\label{cK2}
\eeq
We stress a feature shared by  all
the three functions of $x$ and $y$ in the right-hand side of (\ref{almostf}) taking (\ref{cK}) and (\ref{cK2}) into account.
When $x$ and $y$ are spacelike related, $\sigma(x,y) >0$ so that  the part $2i |t_x-t_y|\epsilon +\epsilon^2$ of  $\sigma(x,y) + 2i |t_x-t_y|\epsilon +\epsilon^2$
does not affect the final result when taking the weak limit $\epsilon \to 0^+$. When $x$ stays in the (causal) future of $y$, $|t_x-t_y| = t_x-t_y$. 
Therefore, for $x$ in the internal region and $y$ in the external region --  so that  (\ref{fundamental}) holds true --
 we can replace $\sigma(x,y) + 2i |t_x-t_y|\epsilon +\epsilon^2$ for the regularized distance appearing in the Hadamard prescription
$\sigma_\epsilon(x,y) = \sigma(x,y) + 2i (t_x-t_y)\epsilon +\epsilon^2$. We can thus write, with $x$ and $y$ as stated above:
$$\langle \Psi, \phi(x)\phi(y) \Psi\rangle  =$$
\beq= A K_0\spa\left(\spa\sqrt{m^2\sigma_\epsilon(x,y)}\right) +
\left[  \frac{m^2}{(2\pi)^2} -  \frac{g^2}{4(2\pi)^2} \left( \frac{1}{2}-\frac{\pi}{3\sqrt{3}} \right) \right]
\spa \frac{K_1\spa\left(\sqrt{m^2\sigma_\epsilon(x,y)}\right) }{\sqrt{m^2\sigma_\epsilon(x,y)}}+
{\cal K}(\sigma_\epsilon(x,y))+\ldots 
\label{final}\eeq
Taking the complex conjugate of both sides of (\ref{final}),  using the fact that $\phi$ is Hermitian, and finally interchanging the name of $x$ and $y$,  from the elementary properties of the $K_\nu$ functions, one easily see that (\ref{final}) holds also for $x$ in the causal past of $y$. {\em So (\ref{final}) holds for all values of $x$ and $y$}.\\

\ssb{ Hawking radiation in the local approach survives the interaction}\label{lastsect}
We intend to analyse the short-distance behaviour of the right-hand side of (\ref{final}) to check if it fits the requirements (R4), especially 
taking Remark \ref{remark} into account. Let us start by considering
 the last term  in the right-hand side of  (\ref{final})
which deserves more attention. 
First of all we recall to the reader  that the function $K_1(\zeta) - 1/\zeta$ is bounded in the closed half-plane $Re \zeta \geq 0$. Boundedness away from the origin easily follows from  8.451(4) and 8.451(6) in \cite{Grad}, while boundedness around the origin is consequence of 
the decomposition: \beq K_1(\zeta) = \frac{1}{\zeta} + I_1(\zeta) \ln(\zeta/2)  + \psi_1(\zeta)\quad \zeta \in \bC \label{expansion}\eeq
where the modified  Bessel function $I_1$ and $\psi_1$ are  holomorphic in the whole complex plane
with  $I_1(0)=0$. The definition of $\cal K$ in (\ref{cK2}) 
yields:
\beq
\cK(\sigma_\epsilon) = 
 \frac{D}{\sigma_\epsilon}
+\frac{h_\epsilon}{\sqrt{\sigma_\epsilon}}\label{agg}\:,
\eeq
where
\begin{eqnarray}
 h_\epsilon(x,y) &:=& \frac{g^2}{4(2\pi)^4}  \int_{2m}^{+\infty}\sp\sp\spa dM \frac{\sqrt{M^2 -4m^2}}{(M^2-m^2)^2}M\left[ K_1(M\sqrt{\sigma_\epsilon}) - \frac{1}{M\sqrt{\sigma_\epsilon}}\right]\:,\label{defh}\\
D&:=&    \frac{g^2}{4(2\pi)^4} \int_{2m}^{+\infty}\sp\sp\spa dM \frac{\sqrt{M^2 -4m^2}}{(M^2-m^2)^2} =
\frac{g^2}{4(2\pi)^2m^2} \left( \frac{1}{2}-\frac{\pi}{3\sqrt{3}} \right)\label{D} \:.
\end{eqnarray}
Since $Re\left(\sqrt{\sigma_\epsilon}\right) \geq 0$, the function in  squared brackets in the integral in the right-hand side of (\ref{defh}) is bounded. Consequently  $h_\epsilon$ satisfies the hypotheses  stated in (1) in Remark \ref{remark}.  
%
The first term  in the right-hand side of  (\ref{final})  can be treated exploiting the known expansion:
\beq K_0(\zeta) =-I_0(\zeta) \ln(\zeta/2)  + \psi_0(\zeta)\quad \zeta\in \bC \nonumber \eeq
where the modified  Bessel function $I_0$ and $\psi_0$ are  holomorphic in the whole complex plane.
The second term  in the right-hand side of  (\ref{final})  can analogously be treated taking advantage of 
 (\ref{expansion}). 
Everywhere $\zeta= m\sqrt{\sigma_\epsilon}$.
It is worth noticing that, due value of $D$ in (\ref{D}), the leading divergence of $\cK(\sigma_\epsilon)$
in (\ref{agg}) exactly cancels an analogous divergence proportional to $g^2$ arising  form the
second term  in the right-hand side of  (\ref{final}).
Collecting all the contributions together, we can easily conclude that 
\beq \omega_2(x,y)  =  \frac{1}{(2\pi)^2\sigma_\epsilon} + \frac{h_\epsilon(x,y)}{\sqrt{\sigma_\epsilon(x,y)}} +  f_\epsilon(x,y)\ln \si_\epsilon(x,y) + r_\epsilon(x,y)\:, \label{ex}\eeq
where the  functions $h_\epsilon, f_\epsilon, r_\epsilon$ verify the conditions stated in (1) in Remark \ref{remark}. Regarding $f_\epsilon$ and $r_\epsilon$, these conditions are fulfilled  because  the Taylor expansions of $I_1$ and $\psi_1$ centred on the origin are  made of odd powers of $\zeta$ only, while those 
of  $I_0$ and $\psi_0$ are  made of even powers of $\zeta$ only.
We stress that  the leading term in the right-hand side of (\ref{ex}) is the same as in the free theory because, as  we have found, the radiative correction give no contribution to the dominant divergence of the two-point function.\\
Since  $\omega_2$ satisfies the requirement (R4), we can apply Theorem \ref{teo1}, obtaining that:\\

\noindent  {\em  Even taking the radiative corrections of the interaction ${\cal L}_I  =\frac{g}{3!}\phi^3$ into account, i.e. referring to the renormalized vacuum state 
$\Psi$ and renormalized field operators $\phi$, at one loop:
$$
\lim_{\la\to 0}|\langle \Psi, \phi(f_\la)\phi(f'_\la) \Psi\rangle|^2 \sim \mbox{C}\:  E_0^2 \: e^{-\beta_H E_0}\:,
$$
for packets sharply concentrated around a large  value $E_0>0$ of the energy when the supports of $f$ and $f'$ are separated by the horizon.
$C$ includes contributions due to the self-interaction.
Finally one also has   the validity of both  (\ref{notunnell}) and the last  statement in Theorem  \ref{teo1} for
 $|\langle \Psi, \phi(f_\la)\phi(f'_\la) \Psi\rangle|^2$,  when  both $f$ and $f'$ have supports in the same region.}

\section*{Acknowledgments.} 
The authors are grateful to R. Brunetti for having pointed out a useful reference and to I. Khavkine for his help in improving the text.\\
This paper partially relies upon G. Collini's Master Thesis in Physics a.y. 2010-2011 (Trento University, supervisors: V. Moretti and N. Pinamonti)

\end{document}